\documentclass{article}\usepackage{natbib,psfig,epsfig,emulateapj}

\newcommand{\beq}{\begin{equation}}
\newcommand{\eeq}{\end{equation}}
\newcommand{\bea}{\begin{eqnarray}}
\newcommand{\eea}{\end{eqnarray}}
\renewcommand{\cite}[1]{\citep{#1}}
\newcommand{\rem}[1]{ }

\bibpunct[,]{(}{)}{;}{a}{}{,}
\begin{document}

\title{Electron acceleration in
relativistic GRB shocks}

\author{Mikhail V. Medvedev\altaffilmark{1}}
\affil{Department of Physics and Astronomy, 
University of Kansas, KS 66045}
\altaffiltext{1}{Also: Institute for Nuclear Fusion, RRC ``Kurchatov
Institute'', Moscow 123182, Russia}

\begin{abstract}
The shock model of gamma-ray bursts (GRBs) contains two equipartition
parameters: the magnetic energy density and the kinetic energy density 
of the electrons relative to the total energy density of 
the shock, $\epsilon_B$ and $\epsilon_e$, respectively. These are free
parameters within the model.
Whereas the Weibel shock theory and numerical simulations fix 
$\epsilon_B$ at the level of $\sim$few$\times(10^{-3}...10^{-4})$, 
no understanding of $\epsilon_e$ exists so far. Here we demonstrate 
that it inevitably follows from the theory that 
$\epsilon_e\simeq\sqrt{\epsilon_B}$. The GRB afteglow data fully agree 
with this theoretical prediction. Our result explains why the electrons
are close to equipartition in GRBs. The $\epsilon_e-\epsilon_B$ relation 
can potentially be used to reduce the number of free parameters in afterglow 
models.
\end{abstract}



\section{Introduction}
\label{s:intro}

Shocks in gamma-ray bursters (GRBs) as well as in many other
astrophysical sources are collisionless. Their physics is much more
complex than that of standard hydrodynamic shocks.
The Weibel shock theory developed by \citet{ML99} has  been confirmed 
by a large number of numerical simulations both in the 
ultra-relativistic regime \cite{Silva+03,Nish+03,Fred+04,Spit05}
and in the non-relativistic one \cite{MSK06}. In particular, 
it has been predicted and then numerically confirmed that the
magnetic equipartition parameter, $\epsilon_B$, defined as
the ratio of the magnetic energy density to the total kinetic 
energy density of a shock,
\beq
\epsilon_B=\frac{B^2/4\pi}{m_pc^2n\Gamma},
\eeq
falls in the range of $\sim$few$\times(10^{-3}...10^{-4})$,
where $n$ is the co-moving particle density and $\Gamma$ is the 
shock Lorentz factor; we assumed $\Gamma\gg1$. 

Multiwavelength spectral fits to afterglows allowed reliable determination
of micro-physics parameters for  a number of GRBs
\cite{PK01a,PK01b,PK02,CL00,LC01,LC03,Yost+03,Frail+03,Berger+04,%
McMahon+04,P05}.
The analyses found that $\epsilon_B$ is indeed falls in the
range $\sim10^{-2}...10^{-5}$, however with relatively large 
errorbars --- typically of about a decade, depending on quality of data 
(afterglow temporal and spectral coverage) and the method of analysis. 
These results confirm 
(within uncertainty) the predictions of the Weibel shock theory. 
We note that this is the only theory, which explains the origin 
of magnetic fields in relativistic shocks.

The afterglow analyses also yield the values of $\epsilon_e$, defined as
the ratio of the kinetic energy density of electrons to the total energy 
density of the shock:\footnote{
Often, $\epsilon_e$ is defined for 
a power-law distributed electrons, so the power-law index enters the 
definition: $\gamma_{e,min}=\left((p-2)/(p-1)\right)(m_p/m_e)\Gamma\epsilon_e$,
where $\gamma_{e,min}$ is the low-energy cut-off of the electron
distribution and $p$ is the power-law index.
In this case, to measure  the total energetics one uses another 
(related) parameter $\epsilon_i$. Here we make no assumption about the 
electron energy distribution, hence our $\epsilon_e$ measures the overall 
energetics of the electrons. Thus our $\epsilon_e$, defined
in Eq. (\ref{ee}), is identical to $\epsilon_i$ used by some authors.
}
\beq
\epsilon_e=\frac{U_e}{m_pc^2n\Gamma}.
\label{ee}
\eeq
The values turn out to be clustered at 
$\sim$few$\times(10^{-1}...10^{-2})$. Given the fact that before being 
shocked the electrons carry only about $\sim m_e/m_p\sim10^{-3}$ 
of the total energy of the ejecta, they must be accelerated somehow  
in the downstream region. Traditionally, one
invokes diffusive Fermi acceleration for this purpose. However,
the Fermi theory cannot accurately predict the value of $\epsilon_e$
in GRB shocks. Moreover, recent studies \cite{NO06,LW06} put the efficiency, 
or even the very presence, of Fermi acceleration at GRB shocks into question.
Thus, until now, we had little understanding of the electron acceleration 
in GRBs and we did not know why the electrons are close to equipartition.

In this paper, we demonstrate that 
$\epsilon_e\simeq\lambda\sqrt{\epsilon_B}$ (with the constant $\lambda\sim1$)
in relativistic, baryon-dominated shocks (i.e., shocks in electron-proton 
plasma). The result is very robust and is based solely on well-known
properties of collisionless shocks.

\section{Evaluation of $\epsilon_{e}$}
\label{s:ee}

Magnetic fields are generated at shocks by the Weibel instability.
In baryon-dominated shocks, the value $\epsilon_B\sim10^{-3}$ is limited by 
the charge-separation effects, which modify the instability 
growth rate and the its dynamics \cite{WA04,Tzon+06}. Thus, saturation 
occurs at the equipartition with the lightest species ---
the electrons. At such low fields, protons keep streaming in current 
filaments,\footnote{
By the way, this explains why numerical simulations
still cannot fully resolve baryon-dominated shocks, in contrast to the 
electron-positron pair-dominated shocks (see, e.g., \citealp{Fred+04,Spit05}).
}
whereas the electrons, being much lighter than the protons, are quickly
isotropized in the random fields and form a uniform background.
The current filaments are formed by the protons moving roughly at 
the speed of light (their Lorentz factor is $\sim\Gamma$). Hence, they  
are sources of both the magnetic and electrostatic fields 
\cite{Hed+04,Nish+05accel}. These fields are related to each other as
\beq
B= \beta E,
\eeq
where $\beta=\sqrt{1-\Gamma^{-1}}\sim1$. An electron, moving toward a
filament gains energy
\beq
u_e\simeq e l E \simeq e l B.
\eeq
The typical radial distance the electron travels is about half the
distance between the filaments, $l\simeq\lambda(c/\omega_{pp,rel})$,
where $\lambda\sim1$ is the dimensionless parameter, $c/\omega_{pp,rel}$
is the relativistic proton skin depth --- the typical scale of structures in
the Weibel turbulence, and $\omega_{pp,rel}=(4\pi e^2 n/m_p \gamma_p)^{1/2}$
is the relativistic proton plasma frequency and $\gamma_p\simeq\Gamma$.
The parameter $\lambda$ accounts for the actual geometry of the filaments,
the electrostatic shielding in plasmas, the effects of the electrons
on the current distribution, etc. All these effects introduce only a small,
factor of two, uncertainty, as is discussed at the end of this section. 
Finally, the electron energy density behind the shock front is
\beq
U_e=n u_e\simeq \lambda eBnc/\omega_{pp,rel}.
\eeq
This equation can be cast into the form 
\beq
\epsilon_e\simeq\lambda\sqrt{\epsilon_B}.
\label{eBee}
\eeq
Note that we didn't make any assumptions here on whether the shock 
compression have already occurred or not (i.e., how far downstream
we are). We just used the fact that the 
shock magnetic fields are due to proton currents, which
also produce electrostatic fields. These electrostatic fields locally
accelerate electrons on their way in and decelerate as they go away 
from the filament. Since the electron emissivity is 
$F_\nu\propto B^2\gamma_e^2$, the electrons strongly radiate
near the filaments, where their energy and the magnetic field are
both at maximum. Hence,
Eq. (\ref{eBee}) represents the emission-weighted relation
between $\epsilon_e$ and $\epsilon_B$. It is this relation that 
should be found in GRB observations.  

In the electron-positron-dominated shocks, the situation is drastically 
different. By symmetry ($e^+$ and $e^-$ have identical masses), the 
current filaments are formed by nearly equal numbers of electrons 
and positrons moving in opposite directions. Hence, the net charge
of such filaments is vanishing and no substantial particle acceleration
is expected. This is in full agreement with simulations \cite{Spit05}. 

A rather important question is: How well is the parameted $\lambda$ 
constrained? We performed a special study using a more detailed 
model of electron dynamics in current filament fields. The details of
this work will be reported elsewhere. Here we just cite the relevant result:
the emission-weighted value of $\lambda$ is very insensitive to the details,
e.g., the current distribution, filament filling factor, etc. For typical
parameters obtained from PIC simulations, $\lambda$ ranges between
$1\la\lambda\la3$ while the filling factor ranges from $\sim1/4$ 
(near the shock front) to $\sim1/100$ (far downstream). 
Thus, $\lambda$ is constrained rather well, it is not just a new free
parameter of a model.

\section{Comparison with observations}
\label{s:obs}

To compare with observations, we have taken the most recent and the
best analyzed sample of data containing ten afterglows \cite{P05}. 
These GRB afterglows were fitted to a number of afterglow models, 
which include combinations of the models of the external media profiles 
(constant density and wind-like) with the models for the ejecta structure
(jetted, structured outflows and energy injection models).
For each model, the reduced $\chi^2$ was given and for models with
``reasonably good fits'' ($\chi^2$/dof$\le4$), the micro-physical 
parameters are given. 

\begin{figure}
\epsscale{.85}
\plottwo{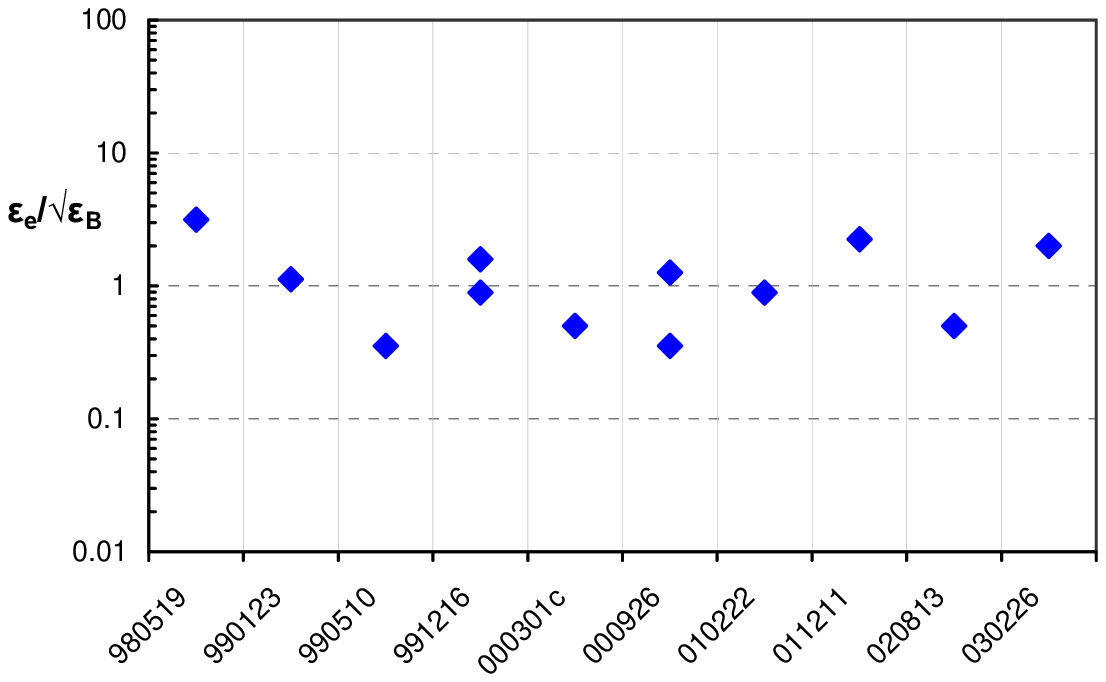}{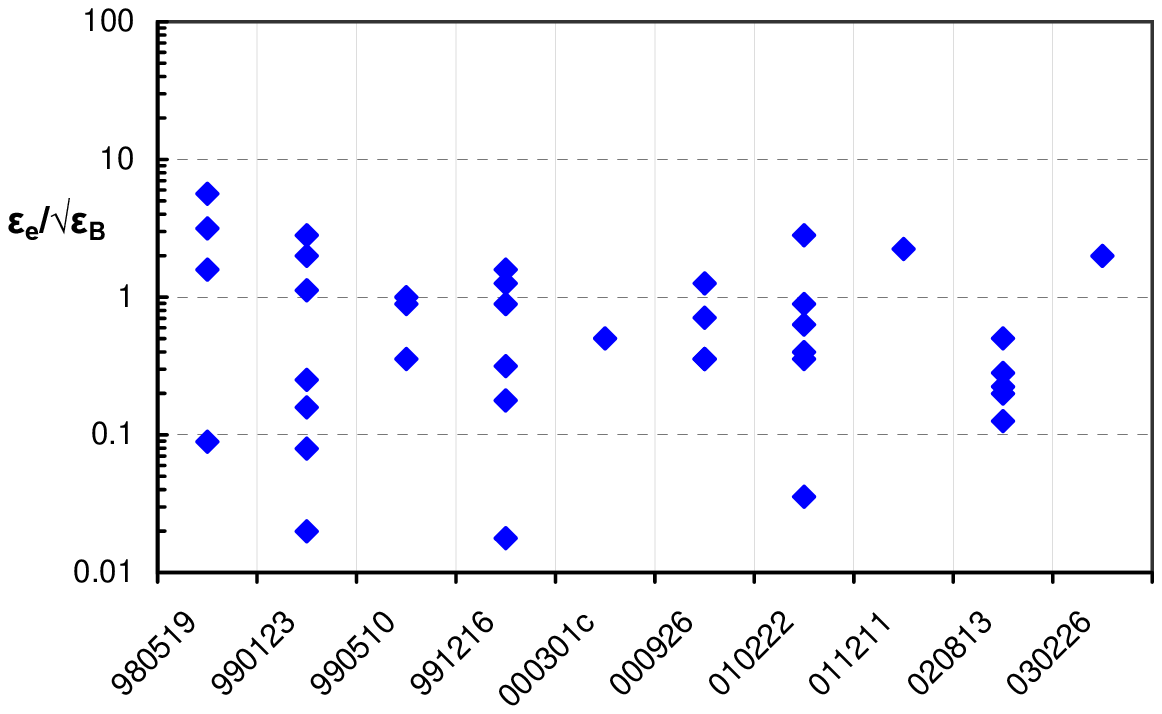}
\caption{The $\epsilon_e/\sqrt{\epsilon_B}$ ratio for ten GRB
afterglows analysed by \citet{P05}. (a) --- The parameters
of the best fit model (with the lowest $\chi^2$/dof)
are used. The clustering of data points around unity is evident. 
All points are consistent with unity within the quoted errorbars.
(b) --- All reported models are used (with $\chi^2$/dof$\le4$). 
The scatter is substantially larger, indicating the degradation 
of the $\epsilon_e-\epsilon_B$ correlation for poor fits. }
\label{f:1} 
\end{figure}

From this data set, we have chosen the best fit model (having the smallest
reduced $\chi^2$ value) for each GRB. The parameters are given in 
Table \ref{t:1}. Note that for two GRBs, there are two equally good fits,
hence we included both. As one can see, $\epsilon_B$ varies over
two orders of magnitude and $\epsilon_e$ (which is $\epsilon_i$ in 
definitions of the paper \citealp{P05}) varies by one order of magnitude.
If $\epsilon_B$ and $\epsilon_e$ are statistically independent, the
scatter of the quantity $\epsilon_e/\sqrt{\epsilon_B}$ should be 
of about two decades. We plot $\epsilon_e/\sqrt{\epsilon_B}$ for 
the best fit models in Figure \ref{f:1}a. Clearly, the data is clustered
around unity with only little scatter. No indvidual confidence intervals
for the best fit parameters were provided in \citet{P05}. Only the 
overall uncertainties were provided: $\sigma(\lg \epsilon_B)=1$,
$\sigma(\epsilon_e)=0.3\epsilon_e$. These yield an uncertainly of the ratio
$\sigma(\lg[\epsilon_e/\sqrt{\epsilon_B}])\simeq0.5$.

The clustering of $\epsilon_e/\sqrt{\epsilon_B}$ near unity is neither 
accidental, not an artifact of fitting. To demonstrate this, we
plot in Figure \ref{f:1}b the values of $\epsilon_e/\sqrt{\epsilon_B}$
for all models reported by \citet{P05}. The goodness of the fit for these
models is $\chi^2$/dof$\le4$. The data points are scattered over almost three
decades, which is consistent with degradation of statistical correlation 
of $\epsilon_e$ and $\epsilon_B$ in poor fits.

Finally, we performed a linear fit in the 
$\log\epsilon_e-\log\epsilon_B$-space. We set the intercept to zero 
($\log\lambda=0$) to reduce the number of degrees of freedom, because 
a two-parameter fit does not give a statistically acceptible result. 
The one-parameter fit yields the exponent in the relation
$\epsilon_e=\epsilon_B^s$ being $s=0.49\pm0.07$ with the $p$-value
$\sim10^{-7}$. This is in excellent agreement with Eq. (\ref{eBee}).

\section{Conclusions}
\label{s:concl}

In this paper, we have shown that in relativistic, baryon-dominated shocks 
$\epsilon_e\simeq\lambda\sqrt{\epsilon_B}$ with the constant $\lambda\sim1$. 
The result inevitably follows from the micro-structure of collisionless
shocks. No {\it ad hoc} assumptions were made. Interestingly, the values of
$\epsilon_e/\sqrt{\epsilon_B}$ derived from afterglow data of ten GRBs 
are clustered around unity, thus supporting our theory.
Since the typical value of $\epsilon_B$ is about $10^{-3}$, the
corresponding value of $\epsilon_e$ should be $\sim0.03$. 
Indeed, the afterglow fits show that 0.03 is the typical value
of the $\epsilon_e$ parameter. Thus, our theory explains why the 
electrons in GRB shocks are close to equipartition. 
Interestingly, the typical uncertainly in $\epsilon_B$ determined
from afterglow fits is often rather large, -- of about an order 
of magnitude or even more. In contrast, the theoretical uncertainly 
in the value of $\lambda$ is only a factor of few at most, and is
very likely even less. Therefore, one can use the obtained relation
$\epsilon_e\sim\sqrt{\epsilon_B}$ to reduce the number of free 
parameters in afterglow models. It will be very interesting 
to investigate how the goodness of afterglow model fits 
changes when the above relation is used. 

A number of important questions are left outside 
of the present study. It would be interesting to calculate the energy 
distribution of electrons. Some simple estimates can, however, be given
as follows. Far away from the filaments, the electrons form the isotropic 
distribution with a typical Lorentz factor $\gamma_{e,min}\sim\Gamma$, because
they have been pitch-angle scattered in small-scale magnetic fields (these
scatterings do not change particle energy). Inside the filaments the
$\gamma$-factor of the electrons shall be maximum, 
$\gamma_{e,max}\sim(m_p/m_e)\Gamma\epsilon_e$. Since these electrons
contribute the most to the observed emission, the emission-weighted 
(that is, ``observed'') distribution will be peaked at $\gamma_{e,max}$,
which shall correspond to $E_{peak}$ in the spectral distribution of GRB 
emission. Other mechanisms are likely needed in order to explain 
the power-law spectra above $E_{peak}$. Numerical PIC simulations 
\citep{Silva+03} indicate that reconnection events occuring during current 
filament coalescences lead to local acceleration of electrons.
This can possibly produce a power-law distribution of electrons.
Alternatively, a power-law radiation spectrum can be produced even by a 
monoenergetic electron distribution, provided the spatial spectrum
of small-scale magnetic fields is a power-law \citep{M06}. PIC simulations
indicate that power-law spectra of magnetic felds indeed form 
at relativistic shocks \citep{Fred+04}. A related question is 
how large the fraction of the energetic electrons is. A simple estimate 
indicates that is should be of order the filling factor of current
filaments, which can be determined from PIC simulations: it varies with 
the downstream distance from $\sim1/few$ to $\sim1/100$  in typical 
electron-proton runs. However, 
more accurate answers to the above questions require detailed modeling
of the electron dynamics, taking into account the residence time of
electrons inside the filaments (the electrons are deflected in the 
direction of the ion current as they move toward a filament). 
We are developing a detailed model and the results will be 
presented in forthcoming publications.

\acknowledgements
This work has been supported by NASA grant NNG-04GM41G 
and DoE grant DE-FG02-04ER54790.

\clearpage

\begin{table}
\caption{\rm 
The best fit afterglow model parameters from \citet{P05}.
The models selected by the lowest reduced $\chi^2$ (fourth column).
Two GRBs, 991216 and 000926, have more than one best fit model.
The models in the last column are SO=structured outflow, 
EI=model with energy injection, JET=model with constant $\Gamma$ within a 
jet opening angle, ISM=constant density medium, WIND=wind-like medium. }
\label{t:1}
\begin{center}
\begin{tabular}{lllll}
\hline\hline
GRB & $\lg\epsilon_B$ & $\lg\epsilon_e$ & $\chi^2$/dof & model \\
\hline
980519  & -3.8 & -1.4 & 1.4  & SO+WIND \\
990123  & -2.1 & -1.0 & 1.5  & EI+WIND \\
990510  & -2.3 & -1.6 & 0.78 & JET+ISM \\
991216  & -3.9 & -2.0 & 1.2  & SO+ISM \\
...     & -3.8 & -1.7 & 1.2  & SO+WIND \\
000301c & -2.6 & -1.6 & 3.3  & SO+ISM \\
000926  & -2.8 & -1.3 & 2.2  & SO+ISM \\
...     & -1.3 & -1.1 & 2.2  & JET+ISM \\
010222  & -3.7 & -1.9 & 1.7  & SO+ISM \\
011211  & -3.3 & -1.3 & 2.3  & SO+ISM \\
020813  & -3.4 & -2.0 & 1.1  & SO+ISM \\
030226  & -3.6 & -1.5 & 4.0  & SO+ISM \\
\hline\hline
\end{tabular}
\end{center}
\end{table}


\end{document}